\documentclass[fleqn,usenatbib]{mnras}

\usepackage{newtxtext,newtxmath}

\usepackage[T1]{fontenc}

\DeclareRobustCommand{\VAN}[3]{#2}
\let\VANthebibliography\thebibliography
\def\thebibliography{\DeclareRobustCommand{\VAN}[3]{##3}\VANthebibliography}

\usepackage{graphicx}	
\usepackage{amsmath}	
\usepackage{bm}
\usepackage{verbatim}

\usepackage{braket}
\usepackage{hyperref}
\usepackage{subfigure}

\title[BTFR of LSBGs]{The baryonic Tully-Fisher relation of HI-bearing low-surface brightness galaxies implies their formation mechanism}

\author[Hua et al.]{
Zichen Hua$^{1,2}$,
Yu Rong$^{1,2}$\thanks{Corresponding author; E-mail: rongyua@ustc.edu.cn},
Huijie Hu$^{3,4}$
\\
$^{1}$Department of Astronomy, University of Science and Technology of China, Hefei 230026, China\\
$^{2}$School of Astronomy and Space Sciences, University of Science and Technology of China, Hefei\\
230026, China\\
$^{3}$University of Chinese Academy of Sciences, Beijing 100049, China\\
$^{4}$National Astronomical Observatories, Chinese Academy of Sciences, Beijing 100012, China
}

\date{Accepted November 28, 2024. Recieved March 25, 2024.}

\pubyear{2024}

\begin{document}
\label{firstpage}
\pagerange{\pageref{firstpage}--\pageref{lastpage}}
\maketitle

\begin{abstract}
     We investigate the baryonic Tully-Fisher relation in low-surface brightness galaxies selected from the Arecibo Legacy Fast ALFA survey. 
     We find that the $\rm HI$-bearing low-surface brightness galaxies still follow the baryonic Tully-Fisher relation of typical late-type galaxies, with a slope of approximately 4 in the baryonic mass versus rotational velocity diagram on the logarithmic scale, i.e., $M_{\rm{b}}\propto v_{\rm{rot}}^4$. 
     Our findings suggest that the matter distributions in low-surface brightness galaxies may resemble that of general late-type galaxies, and hint that low-surface brightness galaxies may not originate from dark matter halos of low densities or stronger/weaker feedback processes, but may emerge from dark matter halos with high spin values. 
\end{abstract}

\begin{keywords}
galaxies: kinematics and dynamics -- galaxies: formation -- galaxies: evolution
\end{keywords}



\section{Introduction}\label{sec_intro}
low-surface brightness galaxies (LSBGs) constitute a special population of galaxies with central surface brightness at least one magnitude fainter than that of the sky background \cite[i.e., $B$-band surface brightness $\rm \geq 22.5 \ mag \cdot arcsec^{-2}$;][]{Impey1997-LSBG, Bothun1997-LSBG}. 
Many isolated LSBGs have been observed to exhibit higher HI fractions (e.g., \citealp{Catinella2018-xGass, Schombert2021-LSBG, Du2015-LSBG, He2020-LSBG}). However, some studies (e.g., \citealp{Martin2019-LSBG, Galaz2002-LSBG}) alternatively found that some LSBGs exhibit the same or even smaller gas fractions compared to high-surface brightness galaxies (HSBGs).
LSBGs usually have low star formation rates (\citealp{Wyder2009-SFR_LSBG, Rong20}), low metallicities (\citealp{Kuzio_de_Naray2004-Oxygen_LSBG, Schombert2021-LSBG}), and tend to appear in low-density environment (\citealp{Perez2019-LSBG_environment, Mo1994-LSBG_spatial, Galaz-LSBG}).
\par

The formation mechanism of LSBGs, which plays a crucial role in refining our empirical galaxy formation models,  have gathered much attention ever since their initial detection. 
Early studies often regarded  LSBGs as galaxies formed in low-density dark matter halos (\citealp{Dekel1986-origin_dwarf, McGaugh1992-LSBG, Mo1994-LSBG_spatial}). However, recent studies have presented more alternative scenarios.
One classical formation scenario is that LSBGs are galaxies hosted in dark matter halos with higher halo spin. In cosmological simulations, the halos of LSBGs are found to have higher spin parameters compared with HSBGs, regardless of whether they are relatively massive (\citealp{Mo1998-disk, Kim2013-LSBG_spin, Kulier2020-LSBG_EAGLE, Perez2022-LSBG_TNG}), or belong to ultra-diffuse galaxies (UDGs, \citealp{Rong2017-UDG, Amorisco2016-UDG, Rong2024-UDG}) -- a specific subset of LSBGs of smaller masses (e.g., \citealp{van_Dokkum_UDG, van2018galaxy, Pina2018-UDG, Pina2019-UDG_BTFR}).
Additionally, some studies suggest that low-mass LSBGs may represent failed $L^{\rm *}$ galaxies (\citealp{van_Dokkum_UDG,van_Dokkum_UDG_spec}), or originate from early co-planar mergers (\citealp{Wright_UDG}) or stellar feedback processes (\citealp{Chan2018-origin_UDG, Dicinto2019-LSBG}), or birth during tidal interactions (\citealp{Rong20b, Carleton19}), or form through multiple pathways (\citealp{Papastergis2017-HI_UDG}).
High-mass LSBGs, alternatively, may form through mergers (\citealp{Saburova-GLSBG_UGC1922, Dicinto2019-LSBG, Zhu2023-GLSBG}), two-stage process with external gas accretions (\citealp{Saburova2021-GLSBG}), or dynamical evolution driven by bars (\citealp{Noguchi2001-GLSBG}).
\par

Nevertheless, despite extensive efforts, our present comprehension of the formation and evolution of LSBGs is still limited.
One of the key reasons is that studies on the mass distributions and baryonic fractions in LSBGs have not achieved a consensus, and a comprehensive scenario has not been established to constrain the current formation and evolution models of LSBGs.
Many studies have shown that the dynamics of LSBGs may be dominated by dark matter (e.g., \citealp{de1997-DM_LSBG, Mowla2017-UDG_tidal, Swaters2003-LSBG, Read2017-MstarMhalo, Pina2022-baryon_halo}), yet UDGs may exhibit a deficiency in dark matter content (e.g., \citealp{van2018galaxy, Pina2019-UDG_BTFR}). 
As for the stellar-to-halo mass relation ($M_{\rm *}-M_{\rm h}$), while \citet{Prole2019-halo} propose that LSBGs form a continuous extension of typical dwarfs in the $M_{\rm *}-M_{\rm h}$ diagram, other studies reveal significant deviations for UDGs from the typical $M_{\rm *}-M_{\rm h}$ relation (e.g., \citealp{van2018galaxy}).
Additionally, some giant LSBGs appear to be dominated by baryons at their central regions, akin to the HSBG counterparts (e.g., \citealp{Lelli2010-gLSBG, Saburova2021-GLSBG, Saburova2019-gLSBG});
the stellar-to-halo mass relation of the giant LSBGs and HSBGs also exhibits a similarity.
All of these confusing signs of diversity in matter distribution among LSBGs could offer valuable insights into their varied formation and evolution pathways. Consequently, a comprehensive understanding of the matter distributions in LSBGs is essential for refining our current models of galaxy formation and evolution.
\par

The baryonic Tully-Fisher relation (BTFR) is a correlation between the rotational velocity, $v_{\rm rot}$, and the baryonic mass, $M_{\rm b}$, of galaxies.  
Over the past two decades, many studies have shown that late-type galaxies and dwarf galaxies consistently adhere to a power-law form of BTFR, despite their wide ranges of baryonic mass, luminosity, and size. BTFR could be written as,
$M_{\rm b} \propto v_{\rm rot}^\beta$,
where the logarithmic slope, $\beta$, typically falls within a range from 3 to 4 (e.g., \citealp{Mcgaugh2000-BTFR, Verheijen2001-TFR, Bugum2008-Dwarf_BTFR, Papastergis2016-BTFR_vary, Lelli2016-small_scatter, Sales2017-BTFR, Ponomareva2018-BTFR, Goddy2023-BTFR}). 
Notably, some studies have also discovered that early-type galaxies may also follow the BTFR of late-type galaxies (\citealp{denHeijer2015-BTFR_ETG, Serra2012-BTFR_ETG}).

In the context of $\rm \Lambda CDM$ paradigm, the presence of BTFR is considered as a correlation between the virial velocity of the dark matter halo and the baryonic mass enclosed within that halo (\citealp{McGaugh2012-BTFR, Mo1998-disk}). Thus, it can serve as a valuable tool to probe the distribution of baryonic matter and dark matter within galaxies.
Furthermore, simulations have highlighted the significance of feedback processes in shaping the observed BTFR (e.g., \citealp{Governato2010-dwarf, Pinotek2011-feedback, Dutton2012-BTFR_outflow}), suggesting that investigating the BTFR can serve as a valuable means to constrain the feedback mechanisms at work within galaxies.
Additionally, BTFR is highly related to other dynamical relationships such as the radial acceleration relation (e.g., \citealp{McGaugh-RAR, Lelli-RAR}). 
Exploring the BTFR could also provide insights into new gravitational theories, as a BTFR with $\beta = 4$ is well predicted by the MOdified Newtonian Dynamics (MOND, \citealp{Milgrom1983-MOND}).
Given its multifaceted implications, the BTFR remains one of the most extensively studied dynamical relations in astrophysics.
\par

As anticipated, recent studies examining the BTFR of LSBGs also show some potential signs of tension. 
Previous studies have found that the 
BTFR does not depend on galactic surface brightness or other galactic properties(e.g., \citealp{Lelli2016-small_scatter, Ponomareva2018-BTFR}; see also \citealp{Zwaan1995-TFR_LSBG}). 
However, some recent investigations (e.g., \citealp{Pina2019-UDG_BTFR, Guo2020-UDG, Hu2023-UDG_BTFR, Karunakaran2020-UDG, Rong2024-UDG}) reveal that, the particular subsample of LSBGs with large disk sizes,  i.e., UDGs, may significantly deviate from the BTFR of typical late-type galaxies, indicative of the possible dependence on size (\citealp{Pina2020-UDG}).
Therefore, a thorough and meticulous examination of the BTFR for LSBGs becomes essential. Such an analysis may not only indicate whether the matter distribution within LSBGs aligns with that of typical late-type and dwarf galaxies, but also provide valuable constraints on the formation mechanisms of LSBGs.
\par

In this work, we will first introduce our sample selection in Section \ref{sec_sample}, and then describe our methods for calculating $M_{\rm b}$ and $v_{\rm rot}$ in Section \ref{sec_method}. 
We will then present our results in Section \ref{sec_result}, and finally discuss the potential selection biases and the possible formation mechanism of LSBGs in Section \ref{sec_discussion}. The summary is in Section \ref{sec_summary}.
\par

\section{Sample Selection}\label{sec_sample}
The Arecibo Legacy Fast ALFA survey (ALFALFA) is a wide-area blind  $\rm HI$ survey aimed at searching for $\rm HI$-bearing objects in the extragalactic neighbourhood. 
\citet{Haynes-a40} matched the $\rm HI$ sources in the ALFALFA $\alpha.40$ catalogue with the optical catalogue of the Sloan Digital Sky Survey (SDSS) Data Release 7 (DR7), and obtained the ALFALFA-SDSS $\alpha.40$ cross-matched catalogue. 
\citet{Du2015-LSBG} and \citet{He2020-LSBG} conducted precise photometry on these cross-matched galaxies in $\alpha.40$ catalog using their SDSS-DR7 $g$- and $r$-band images, obtaining the central surface brightness in the $g$- and $r$-bands, $\mu_0 (g)$ and $\mu_0 (r)$. The $g$- and $r$-band central surface brightness were further converted into $B$-band surface brightness based on (\citealp{Smith2002-ugirz}), 
\begin{equation}
    \mu_0(B) = \mu_0 (g) + 0.47\times[\mu_0(g) - \mu_0(r)] +0.17,
\end{equation}
to select LSBGs. The authors finally selected a sample of LSBGs with the $B$-band apparent central surface brightness of $22.5 < \mu_{0}(B) < 27.64 \ {\rm mag/arcsec^2}$, which is treated as the parent sample and used in this work to study the BTFR of LSBGs.
 
\par

\begin{figure*}
    \centering
    \subfigure{
        \includegraphics[width=0.33\textwidth]{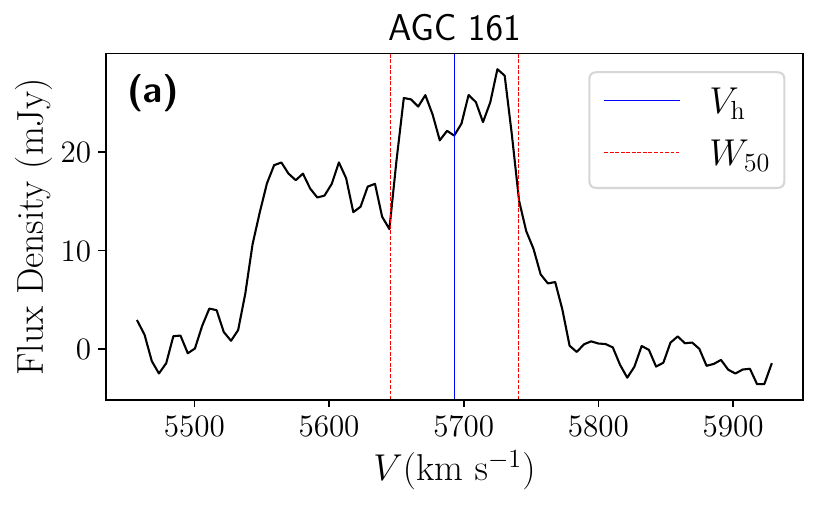}
    }\subfigure{
        \includegraphics[width=0.33\textwidth]{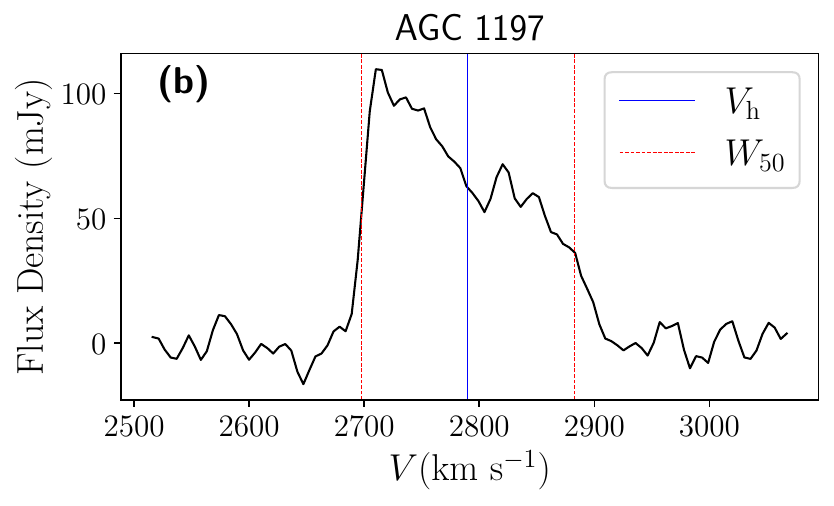}
    }\subfigure{
        \includegraphics[width=0.33\textwidth]{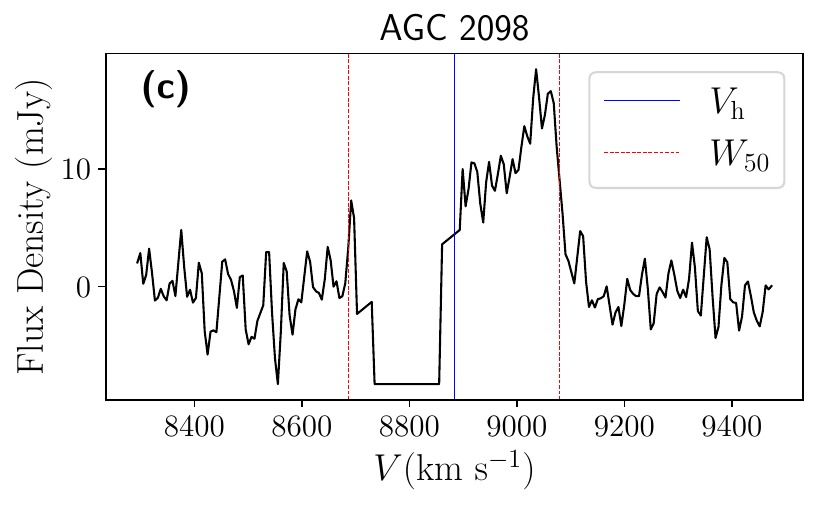}
    }
    \caption{Examples of suspicious HI spectra from \protect{\citet{Haynes2018-ALFALFA}}. $V_{\rm h}$ denotes the heliocentric velocity, and $W_{\rm 50}$ denotes the 50\% peak width of an HI spectrum, obtained by \protect{\citet{Haynes2018-ALFALFA}}. Panel (a): a profile showing potential interactions with other galaxies. Panel (b): a seriously asymmetric profile. Panel (c): a profile containing bad points due to data quality.}
    \label{fig_HI}
\end{figure*}

From the parent sample, we first remove LSBGs with $g$-band apparent axis ratios $b/a > 0.72$, approximately corresponding to the threshold of inclination angles of $i \lesssim 45^\circ$, because the face-on galaxies may introduce large uncertainties in the measurements of the rotation velocities (see equation (\ref{eqs_v_20})). 
We then exclude LSBGs with low $\rm HI$ spectral signal-to-noise ratios ${\rm SNR} < 10$ (see \citealp{Haynes2018-ALFALFA} for the definition of SNR). 
Furthermore, we exclude LSBGs with suspicious HI profiles that indicate potential interactions with other galaxies, serious asymmetric profiles, or profiles with data points affected by poor data quality, as illustrated in Fig.~\ref{fig_HI}. The first two scenarios may suggest an in-equilibrium state of the galaxy, leading to inaccurate estimation of the rotation velocity.
\par

Due to the bad spatial resolution of ALFALFA, 
if a galaxy has a close companion galaxy, its $\rm HI$ spectrum may be contaminated by the companion. To eliminate this possibility, we remove the galaxies with companions within the projected radii of $3.8'$ (i.e., approximately the beam size of Arecibo) and radial velocity differences of $\Delta v<500 {\rm km/s}$.
The redshift of a LSBG is estimated from the central velocity of its $\rm HI$ spectrum
\footnote{While the distances of our sample galaxies are obtained directly from \citealp{Haynes2018-ALFALFA}, please refer to their work for more details.
}, while the redshift of the neighbouring galaxies comes from the SpecObj and Photoz databases of SDSS. Finally, 204 LSBGs are left in our sample.
\par

We also calculate the kurtosis coefficient, $k_{\rm 4}$, of the HI spectrum (\citealp{Papastergis2016-BTFR_vary,Badry2018-HI}) for each galaxy, as
\begin{equation}
    k_4 = M_4/(\sigma^2)^2 - 3, 
\end{equation}
where $M_4$ and $\sigma^2$ are calculated as,
\begin{align}
    M_{\rm 4} &= \int^{\nu_{\rm max}}_{\nu_{\rm min}} (\nu-\bar{\nu})^4 F(\nu)d \nu / \int^{\nu_{\rm max}}_{\nu_{\rm min}} F(\nu)d \nu , \label{eqs_M4} \\
    \sigma^2 &= \int^{\nu_{\rm max}}_{\nu_{\rm min}} (\nu-\bar{\nu})^2 F(\nu)d \nu / \int^{\nu_{\rm max}}_{\nu_{\rm min}} F(\nu)d \nu. \label{eqs_s2}
\end{align}
Here, $\nu$ and $F(\nu)$ represent the frequency and corresponding flux density, respectively. $\bar{\nu}$ is defined as, 
\begin{equation}
    \bar{\nu} = \int^{\nu_{\rm max}}_{\nu_{\rm min}} \nu F(\nu)d \nu / \int^{\nu_{\rm max}}_{\nu_{\rm min}} F(\nu)d \nu.
\end{equation}
where $\nu_{\rm min}$ and $\nu_{\rm max}$ correspond to the minimal and maximum frequencies between which $W_{20}$ (see Section \ref{sec_method} for the definition of $W_{20}$) is measured, respectively. 
$k_{\rm 4}$ can distinguish the single-horned and double-horned spectra (see, e.g., Fig. 1 of \citealp{Badry2018-HI}). The galaxies with single-horned HI lines may have kinematics strongly influenced by velocity dispersion (\citealp{Badry2018-HI}) or beam smearing, or HI disks that are not extended enough to trace the flat part of the rotation curve (\citealp{Papastergis2016-BTFR_vary}). As a consequence, their HI-widths cannot represent the rotation velocities (\citealp{McGaugh2012-BTFR, Verheijen1997-TFR}). Therefore, we perform a cut at the HI spectrum, $k_{\rm 4} < -1$, in order to remove the galaxies with single-horned HI lines (see Fig. 3 of \citealp{Badry2018-HI} for the threshold). 
\par

\begin{figure}
    \centering
    \subfigure{
    \includegraphics[width = 0.85\columnwidth]{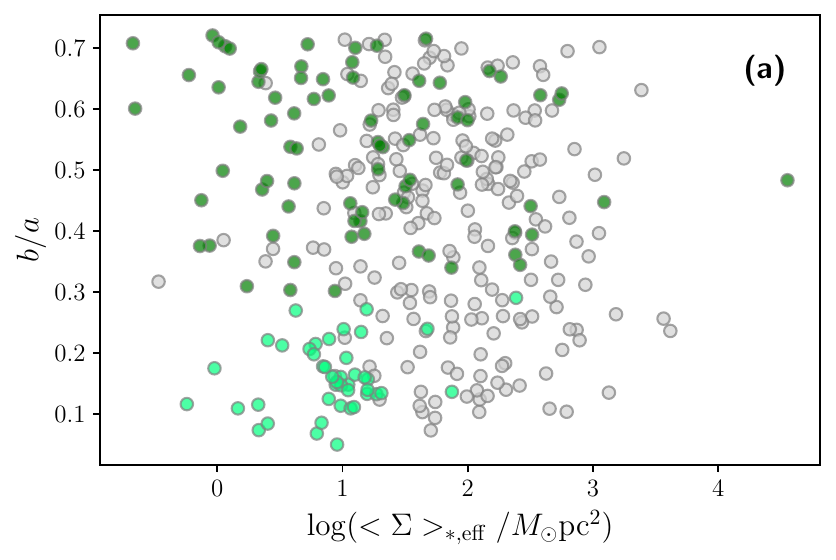}
    }
    \subfigure{
    \includegraphics[width = 0.85\columnwidth]{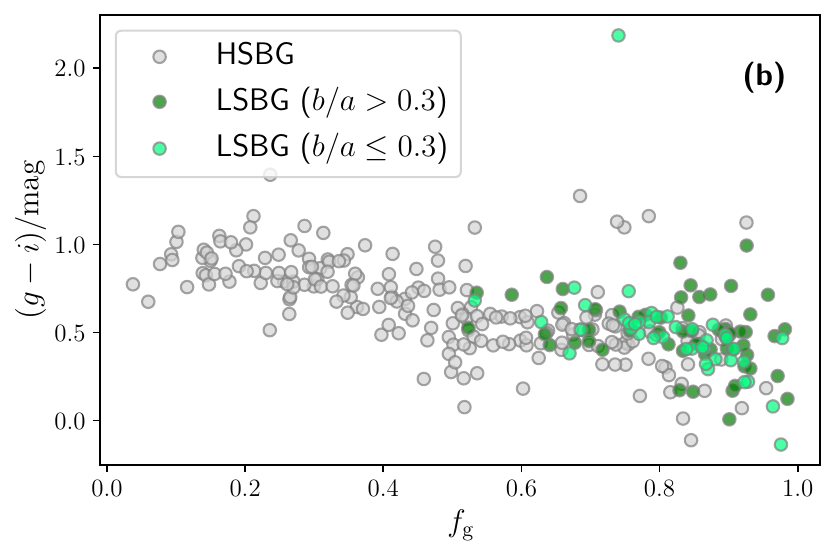}
    }
    \subfigure{
    \includegraphics[width=0.85\columnwidth]{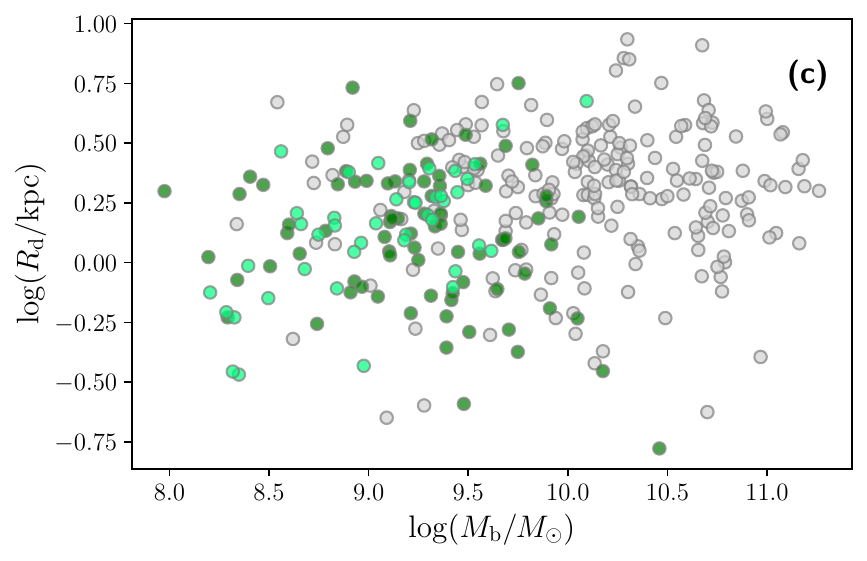}
    }
    \subfigure{
    \includegraphics[width=0.85\columnwidth]{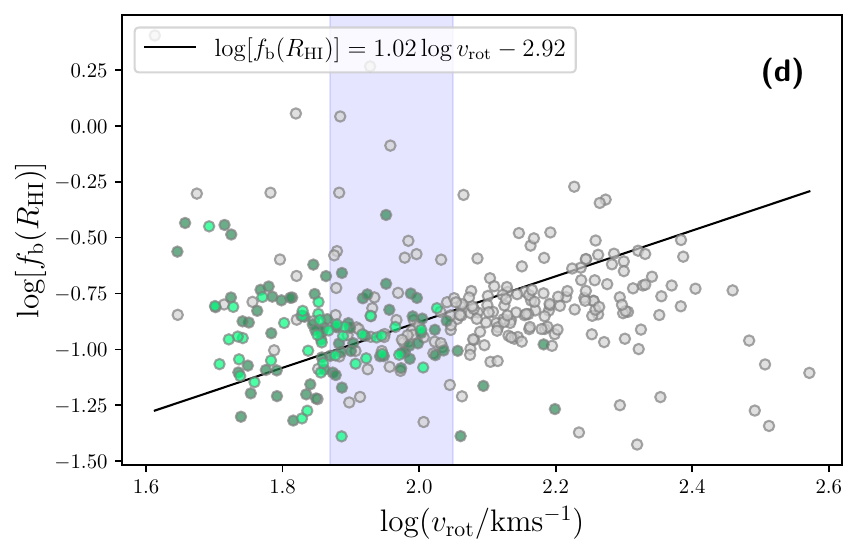}
    }
    \caption{A comparison of properties of selected LSBGs and HSBGs. The green-filled circles represent LSBGs (the dark-green and light-green depict the galaxies with apparent axis ratios $b/a>0.3$ and $b/a\leq 0.3$, respectively), while the grey-filled circles represent HSBGs.
    Panel~(a): axis ratio $b/a$ versus mean surface density $\braket{\Sigma}_{*,\rm{eff}}$. Panel~(b): $g-i$ color versus gas fraction $f_{\rm{g}}$. Panel~(c): scale length $R_{\rm{d}}$ versus baryonic mass $M_{\rm{b}}$. Panel (d): $v_{\rm rot}$ versus $f_{\rm b}(R_{\rm HI})$. Please refer to Section \ref{sec_sample} and Section \ref{sec_discuss_implication} for the definitions of these parameters. The black-solid line and the blue-shaded region in panel~d represent the best-fitting linear line and the bin used for performing the two-sample Kuiper tests, respectively, as described in Section \ref{sec_discuss_implication}.
    }
    \label{fig_Prop}
\end{figure}

The final sample contains 124 LSBGs with double-horned HI lines, ensuring the most reliable results. For comparison, 210 HSBGs with $\mu_{\rm 0}(B) < 22.5 \ {\rm mag}$, and double-horned HI lines are selected from $\alpha.40$ as the counterparts.
Fig.~\ref{fig_Prop} presents various properties of our selected LSBGs and HSBGs, including $g$-band axis ratio $b/a$, average stellar-mass surface density within the $g$-band half-light radius $\braket{\Sigma}_{\rm *, eff} = M_{\rm *}/(2 \pi R^2_{\rm eff})$ (see Section~\ref{sec_method} for the stellar mass $M_{\rm *}$ estimation; $R_{\rm eff}$ is estimated following \citet{Rong2024-UDG}), $g-i$ color, gas fraction $f_{\rm g}$ ($f_{\rm g} = M_{\rm g}/M_{\rm b}$; see Section~\ref{sec_method} for the derivation of $M_{\rm g}$), $g$-band stellar disk scale length $R_{\rm d} = R_{\rm eff}/1.678$, and baryonic mass $M_{\rm b}$.

\par

\section{Method}\label{sec_method}
The baryonic mass of a galaxy is mainly composed of the stellar mass and gas mass.
The stellar mass $M_{\rm *}$ can be estimated with the galaxy luminosity in the $W1$-band (denoted as $L_{ W1}$) from the Wide-field Infrared Survey Explorer (\citealp{2010WISE}), using the method proposed by \citet{McGaugh2015-BTFR}, as
\begin{align}\label{eqs_Ms}
    M_{\rm *}/M_{\rm \odot} = 0.45 L_{ W1}/L_{\rm \odot}.
\end{align}
As for the gas mass, $M_{\rm g}$, we estimate it by adding up the masses of HI and helium, where $M_{\rm HI}$ is the HI mass (see \citealp{Haynes2018-ALFALFA} for the method of estimating $M_{\rm HI}$). The baryonic mass of a galaxy  is then estimated as
    $M_{\rm b} = M_{\rm *} + M_{\rm g}$.
Note that $M_{\rm b}$ does not include the hot ionized gas and molecular gas, as their fractions are expected to be small compared to the HI fraction (\citealp{McGaugh2012-BTFR, Zhong2023-hot_gas}), and thus are negligible.
\par

We use the $20\%$ peak width of an HI spectrum (\citealp[$W_{\rm 20}$;][]{Guo2020-UDG, Hu2023-UDG_BTFR, Rong2024-UDG}; see the \textit{Method} section of \cite{Guo2020-UDG} for the method of estimation of $W_{\rm 20}$)to calculate the rotation velocity $v_{\rm rot}$. \cite{Guo2020-UDG} have demonstrated that $W_{\rm 20}$ can accurately trace galactic $v_{\rm rot}$, which can be estimated as,
\begin{equation}\label{eqs_v_20}
    v_{\rm rot} = \frac{W_{20}}{2 \sin i},
\end{equation}
where $i$ is the inclination angle. $W_{20}$ employed in equation~({\ref{eqs_v_20}}) has been corrected for instrumental broadening and redshift (\citealp{Guo2020-UDG, Catinella2012-GALEX, Kent2008-ALFALFA}).
The errors of $W_{\rm 20}$ are estimated following the method described by \citet{Guo2020-UDG}. Please refer to the \textit{Method} section of their paper for details.
In equation~(\ref{eqs_v_20}), we assume an alignment between the inclination angles of the stellar disk and HI disk in a galaxy and then use the inclination angle of the stellar disk to estimate the rotation velocity. The inclination is estimated from the $g-$band apparent axis ratio, as (e.g., \citealp{Guo2020-UDG, Hu2023-UDG_BTFR, Du2019-LSBG, Rong2024-UDG, Bugum2008-Dwarf_BTFR}), 
$\sin i = \sqrt{\frac{1-(b/a)^2}{1-q_0^2}}$, 
where $q_0$ denotes the intrinsic axis ratio seen edge-on. We take $q_0 = 0.2$, which is commonly used in the previous studies (e.g., \citealp{Du2019-LSBG, Guo2020-UDG, Giovanelli1997-TFR, Tully2009-distance}).
For those edge-on galaxies with $b/a \leq q_0$, we set $\sin i = 1$. 

Previous studies have revealed that the stellar disk and gas disk in galaxies may not be perfectly co-planar, but often exhibit a small inclination difference of ${\rm \delta}i < 20^\circ$ (e.g., \citealp{Starkenburg-counterrotation, Guo2020-UDG, Gault2021-VLA_UDG}). In order to take this misalignment into account, we assume ${\rm \delta}i$ following a Gaussian distribution centred at $0^\circ$ with a standard deviation of $\sigma_i = 20^\circ$. We treat $\sigma_i = 20^\circ$ as the uncertainty associated with $i$ in our study.
Note that we do not apply corrections for the velocity dispersion in equation~(\ref{eqs_v_20}), since the effects of the dispersion are severe under $v_{\rm rot} \lesssim 40 \ {\rm km/s}$ (summarized by \citealp{Lelli2022-gas_review}) while $v_{\rm rot}$ of our sample galaxies are well above $ \sim 50 \ {\rm km/s}$.
\par

\section{BTFR}\label{sec_result}

\begin{figure}
    \centering
    \includegraphics*[width=1.0\columnwidth]{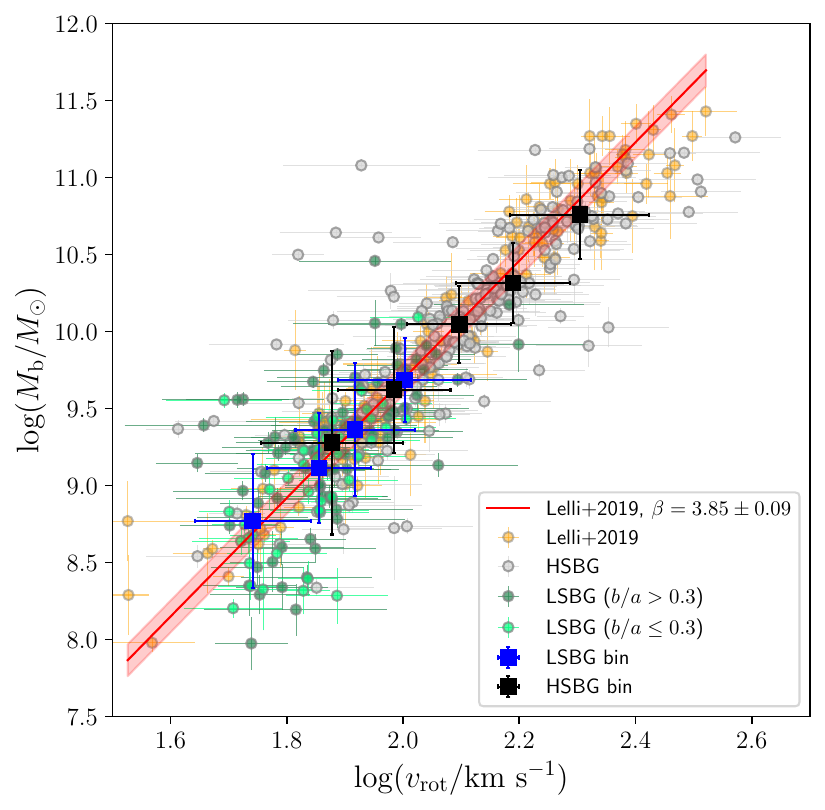}
    \caption{The $v_{\rm rot}-M_{\rm b}$ diagram for our samples.
    The green- and the grey-filled circles are sthe same as those in Figure~\ref{fig_Prop}.
    The orange-filled circles are SPARC galaxies (\protect\citealp{Lelli2016-SPARC}) from \protect\citet{Lelli2019-BTFR}. All the error bars denote $\rm 1\sigma$ uncertainties.
    The blue squares represent the median values of LSBGs while the black squares represent those of HSBGs.
    The red-solid line is the tightest BTFR reported by \protect{\citet{Lelli2019-BTFR}}. 
    The red-shaded area is the $\rm 1\sigma$ intrinsic scatter of the red-solid line, which is considered as the $\rm 1\sigma$ uncertainty of the tightest BTFR.
    }
    \label{BTFR}
\end{figure}

In Fig.~\ref{BTFR}, we observe that the distributions of LSBGs and HSBGs exhibit no significant deviation in the $v_{\rm rot}-M_{\rm b}$ diagram. 
We impose a power-law equation,
\begin{equation}\label{eqs_BTFR}
    \log(M_{\rm b}) = \alpha + \beta  \log(v_{\rm rot}),
\end{equation}
where $\alpha$ and $\beta$ are free parameters to fit the BTFRs of the HSBGs and LSBGs using \textit{scipy.odr} package, respectively.
To estimate the errors of $\alpha$ and $\beta$, we apply the Monte Carlo method. We generate data samples based on the original ($v_{\rm rot}$, $M_{\rm b}$) values and their associated uncertainties. We assume Gaussian distributions for $v_{\rm rot}$ and $M_{\rm b}$ centered at their origins, with the $1\sigma$ regions matching the error bars. This process yields a set of ($\alpha$, $\beta$) values.
We repeat this sampling procedure 5,000 times to obtain 5,000 sets of ($\alpha$, $\beta$) values. The standard deviations of these parameters ($\sigma_{x, {\rm MC}}$) are then combined with the fitting errors ($\sigma_{x, {\rm odr}}$) to determine the uncertainties of $\alpha$ and $\beta$, as
\begin{equation}
    \sigma_{x} = \sqrt{\sigma_{x, {\rm MC}}^2 + \sigma_{x, {\rm odr}}^2}, 
\end{equation}
where the subscript $x$ denotes $\alpha$ or $\beta$.

The best-fitting parameters of the BTFR are  $\alpha_{\rm HSBG} = 1.83 \pm 0.51$ and $\beta_{\rm HSBG} = 3.91 \pm 0.24$ for HSBGs, while are $\alpha_{\rm LSBG} = -1.45 \pm 1.49$ and $\beta_{\rm LSBG} = 5.67 \pm 0.78$ for LSBGs.
\par

We further split LSBGs and HSBGs into the different $\log v_{\rm{rot}}$ bins. Within each bin, we calculate the median value of $\log M_{\rm{b}}$ and the corresponding $1\sigma$ standard deviation for both samples. 
As shown in Fig.~\ref{BTFR}, the binned points of LSBGs (depicted by the blue squares) and HSBGs (depicted as the black squares) comfortably lie within the $1\sigma$ uncertainty region of the tightest BTFR of typical late-type galaxies reported by \citet{Lelli2019-BTFR} (the red-solid line and the red-shaded region denote the best-fitting BTFR of typical late-type galaxies and its $1\sigma$ uncertainty, respectively, as shown in Fig.~\ref{BTFR}), with $\beta\simeq 3.85 \pm 0.09$.
\par

Therefore, we conclude that the BTFRs of LSBGs and HSBGs show no significant deviation. Both galaxy samples follow the BTFR of typical late-type galaxies, with a BTFR slope of approximately 4. 
\par

Our results are consistent with many previous observational (e.g., \citealp{McGaugh2012-BTFR, Lelli2016-small_scatter, Papastergis2016-BTFR_vary}) and simulation (e.g., \citealp{Dutton2017-NIHAO}) results. 
However, it is also worth noting that some other studies alternatively report smaller slopes ($\beta \sim 3$, e.g., \citealp{Goddy2023-BTFR, Ponomareva2018-BTFR}).
\par

\section{Discussion}\label{sec_discussion}
\subsection{Potential biases}
In our study, the SNR threshold used to select the LSBG and HSBG samples may exclude some galaxies with similar HI masses but wider HI line widths, potentially introducing deviations in the slope of the BTFR. To investigate the possible selection effects, we adjust the SNR threshold by increasing it from 10 to 15, and then to 20. We find that, for HSBGs, the best-fitting $\beta$ of their BTFR changes to $4.05 \pm 0.21$ for SNR $> 15$, and $4.05 \pm 0.26$ for SNR $> 20$, still in good agreement with the best-fitting $\beta$ value for SNR $> 10$. For LSBGs, the best-fitting $\beta$ becomes $\sim 6.04 \pm 0.66$ (SNR $> 15$) and $\sim 6.13 \pm 0.82$ (SNR $> 20$), which is also similar to the $\beta \sim 5.67 \pm 0.78$ when applying the selection criterion of SNR $> 10$, considering the large uncertainties of the $\beta$ values. Additionally, the binned points of LSBGs and HSBGs within different $\log v_{\rm{rot}}$ bins for the different SNR thresholds are always located within the $1\sigma$ uncertainty region of the BTFR of typical late-type galaxies. Consequently, we conclude that, for both the LSBG and HSBG samples, the selection effect introduced by the SNR threshold is negligible. 
It should also be noted that, due to the sensitivity limitations of the Arecibo telescope, galaxies with higher rotational velocities $v_{\rm rot}$ exhibit lower signal-to-noise ratios in their HI spectra under a fixed total HI flux, leading to a bias in our parent sample (\citealp{Haynes-a40, Dou2024-HI}).
\par

Another potential bias may arise from the $k_4$ cut. Some rotation-supported galaxies with small $v_{\rm rot}$ may exhibit single-horned spectra due to beam-smearing effects (\citealp{Swaters2003-LSBG, Teodoro2015-BAROLO}), leading to their exclusion by the $k_4$ cut. Additionally, galaxies with larger inclination angles tend to have smaller $k_4$ values (\citealp{Badry2018-HI}). We study the effect of the $k_4$ cut by adjusting the $k_4$ threshold from -1.0 to -1.2, and then from -1.0 to -0.8. The BTFR slope of LSBGs changes from $\sim 5.67 \pm 0.78$ ($k_4 < -1.0$) to $\sim 5.38 \pm 0.70$ ($k_4 < -1.2$) and $\sim 4.28 \pm 0.32$ ($k_4 < -0.8$), while the slope of HSBGs changes from $\sim 3.91 \pm 0.24$ ($k_4 < -1.0$) to $\sim 3.78 \pm 0.24$ ($k_4 < -1.2$) and $\sim 3.85 \pm 0.16$ ($k_4 < -0.8$), respectively. It is worth noting that when the $k_4$ threshold exceeds $-0.8$, some galaxies appear to deviate from the BTFR. However, these galaxies may be strongly dominated by velocity dispersions, as they typically have unreliable $v_{\rm rot} \lesssim 40 \ {\rm km/s}$ (\citealp{Lelli2022-gas_review, Badry2018-HI}). Since we cannot determine whether the single-horned HI spectra of these galaxies are due to the beam-smearing effect or the dominance of the velocity dispersions, we keep $k_4 < -1$ to exclude these galaxies.

\par

Additionally, possible bias emerges from using $W_{20}$ or $W_{50}$ (the $50\%$ peak width of an HI spectrum) to estimate the rotation velocities. However, we find that the average difference between the $\log v_{\rm rot}$ estimated from $W_{20}$ and $W_{50}$ is quite small, even smaller than the uncertainties of $\log v_{\rm rot}$. Therefore, using $W_{20}$ or $W_{50}$ would not significantly affect our results. Besides, we also test whether using the different methods of estimating $M_{\rm *}$ (\citealp{Taylor2011-Mstar}) or considering the intrinsic thickness $q_0$ dependence on $M_{\rm *}$ (\citealp{Lelli2016-CDR}) would lead to the changes of BTFR slopes of LSBGs. We observe no strong changes in BTFR slopes in these tests. Furthermore, we notice that the surface brightness limit determined by the optical survey may lead to large uncertainties in $b/a$ estimation for LSBGs (\citealp{Pina2024-UDG}). However, the large uncertainties should result in a larger scatter, rather than a systematic bias or a change of BTFR slope.
\par

\subsection{Implications on the formation mechanisms of LSBGs}\label{sec_discuss_implication}

Previous theoretical studies posit that LSBGs may form in low-density dark matter halos (e.g., \citealp{Dekel1986-origin_dwarf, McGaugh1992-LSBG}). 
However, this model predicts a significant difference in the BTFRs between LSBGs and HSBGs (\citealp{McGaugh2015-LSBG_formation}), which contradicts our findings.
Therefore, the formation model based on low-density dark matter halos may not refer to LSBGs.
\par

Many observational studies also suggest that LSBGs might be dominated by dark matter compared with HSBGs (e.g., \citealp{de1997-DM_LSBG, Swaters2003-LSBG, Pina2022-baryon_halo, Read2017-MstarMhalo}).
Therefore, we delve deeper into understanding the fraction of dark matter in LSBGs. Note that BTFR could be expressed as 
\footnote{Equation (\protect{\ref{eqs_BTFR_fb}}) is from the \href{https://www.lellifederico.com/_files/ugd/8da6c8_ec117ff185a64da59543dfe118602110.pdf}{lecture on the Tully-Fisher relation}; see also \citet{Aaronson1979-distance} and \citet{Zwaan1995-TFR_LSBG}.},
\begin{equation}\label{eqs_BTFR_fb}
    M_{\rm b} \simeq \frac{f_{\rm b}^2(R)}{Gg_{\rm bar}(R)}v_{\rm rot}^4.
\end{equation}
Here, the baryonic mass fraction within a radius $R$ is defined as $f_{\rm b}(R) = M_{\rm b}/M_{\rm tot}(R)$,
where $M_{\rm{tot}}(R)$ denotes the total mass enclosed within $R$.
$g_{\rm bar}(R)$ is the baryonic acceleration at $R$. 
$g_{\rm bar}(R)$ is calculated by assuming the stellar and gas components in a galaxy follow the razor-thin disk distribution, using the method outlined in \citet{Read2017-MstarMhalo} (see  equation (5) of \citet{Read2017-MstarMhalo} for details).

Note that equation~(\ref{eqs_BTFR_fb}) holds true only when $R$ is large. Here, we utilize the HI radius, $R_{\rm HI}$, defined as the radius where the HI surface density equals $1 \ M_{\rm \odot}/{\rm pc^2}$.
$R_{\rm HI}$ can be estimated based on the strong correlation between $R_{\rm HI}$ and $M_{\rm HI}$ (\citealp{WangJ-2016, Gault2021-VLA_UDG, Lutz2018-HIx}), 
\begin{equation} 
\log\left(\frac{2R_{\rm HI}}{\rm kpc}\right) = (0.506 \pm 0.003) \log\left(\frac{M_{\rm HI}}{M_{\rm \odot}}\right) - (3.293 \pm 0.009). 
\end{equation}
Consequently, we calculate $f_{\rm b}(R_{\rm HI})$ from equation~(\ref{eqs_BTFR_fb}) and then analyze the $f_{\rm b}(R_{\rm HI})\--v_{\rm rot}$ relationship for the LSBGs and HSBGs.

As shown in panel (d) of Figure~\ref{fig_Prop}, in statistics, $f_{\rm b}(R_{\rm HI})$ increases with rising $v_{\rm{rot}}$, with a slope of $\sim 1.02$ using a bisector fitting (\citealp{Isobe-OLS}). 
Since the dark matter halo mass ($M_{\rm h}$) of a galaxy is tightly related to $v_{\rm{rot}}$, as $M_{\rm{h}}\propto v_{\rm{rot}}^3$ \citep{Mo1998-disk}, the correlation between $f_{\rm b}(R_{\rm HI})$ and $v_{\rm{rot}}$ hints at a positive association between the baryonic mass fractions and dark matter halo masses for these galaxies (\citealp{Ayromlou2023-DM, Duffy2010-DM_halo}).
Similar trends have also been observed based on spatially resolved data (e.g., \citealp{Lelli-RAR, Posti2019-scaling, Pina2022-baryon_halo, Teodoro2023-scaling}).
This correlation does not exceed our expectations, given that our galaxy samples are dominated by the low-mass galaxies. 
Halos with lower $M_{\rm{h}}$ have shallower potentials (\citealp{Duffy2010-DM_halo}), making it more difficult to prevent the galactic interstellar medium (ISM) against expulsion by the feedback processes in these shallower potentials, which leads to lower $f_{\rm b}(R_{\rm{HI}})$ (e.g., \citealp{Bryan2013-DM, Duffy2010-DM_halo}). 

Subsequently, in order to compare the $f_{\rm b}(R_{\rm{HI}})$ of the LSBG and HSBG samples with similar dark matter halo masses, we further focus on a specific $v_{\rm{rot}}$ bin defined by $\log v_{\rm rot} \in [1.87,2.05]$ (as depicted by the blue-shaded region in panel (d) of Fig.~\ref{fig_Prop}). 
We have 52 LSBGs and 52 HSBGs in this small bin, exhibiting similar distributions of $v_{\rm rot}$ with median $\log v_{\rm rot}$ values of $1.94 \pm 0.05$ (for LSBGs) and $1.97 \pm 0.05$ (for HSBGs). 
The Kuiper test between the distributions of $v_{\rm rot}$ of LSBGs and HSBGs in this small bin also gives a large $p$-value of $0.77$, indicating no significant difference in $v_{\rm rot}$. 
We then compare the $f_{\rm b}(R_{\rm HI})$ distributions of the LSBG and HSBG sample galaxies within this $v_{\rm rot}$ bin. 
The median values of the $f_{\rm b}(R_{\rm HI})$ distributions of LSBGs and HSBGs are $-0.94 \pm 0.15$ and $-0.91 \pm 0.30$, respectively; the Kuiper test for the two $f_{\rm b}(R_{\rm HI})$ distributions also yields a large $p$-value of $0.34$. 

Therefore, we note that, for LSBGs and HSBGs with similar halo masses, their baryonic fractions are also comparable. 
LSBGs may not be particularly dark matter-dominated galaxies compared to their HSBG counterparts; at least within the regimes of $R_{\rm HI}$, the dark matter fractions of LSBGs and HSBGs are similar.
Furthermore, this result also plausibly suggests comparable feedback power among the LSBGs and HSBGs with similar halo masses (i.e., similar gravitational potentials). 
Consequently, we conclude that feedback may not be the primary driver of the different surface brightness between these HI-bearing galaxies. 
In other words, these HI-bearing LSBGs may not originate from stronger or weaker feedback.

Alternatively, our finding aligns with the simulation results where the LSBGs can be well reproduced by higher dark matter halo spins (e.g., \citealp{Perez2022-LSBG_TNG, Dicinto2019-LSBG}). 
Therefore, we conclude that our result may lend support to the theorem that LSBGs may originate from high-spin dark matter halos (e.g., \citealp{Mo1998-disk}). 
\par

Except for the aforementioned formation models of LSBGs in the framework of $\rm \Lambda CDM$, MOND may also offer a reasonable explanation for the BTFR of LSBGs (e.g., \citealp{Milgrom1983-MOND, Wittenburg-form_MOND}).
As demonstrated in Section \ref{sec_result}, both LSBGs and HSBGs follow a BTFR with a slope indistinguishable from 4. This result adheres to the theoretical predictions by MOND (\citealp{Milgrom1983-MOND}). Therefore, our result cannot exclude the MOND theory.
Nevertheless, several studies on the dynamics of galaxies have produced results that deviate from MOND's predictions (e.g., \citealp{Ren2019-SIDM, Mercado2023-RAR, Pina2024-UDG, Khelashvili2024-SPARC}). Therefore, more compelling studies are needed to better determine if MOND is a promising framework.
\par

\section{Summary}\label{sec_summary}
Taking advantage of the ALFALFA HI survey data, we have selected the HI-bearing LSBGs and HSBGs, and estimated their baryonic masses $M_{\rm b}$ and rotation velocities $v_{\rm rot}$.
Our findings reveal that, statistically, the HI-bearing LSBGs and HSBGs also conform to the BTFR found in typical late-type galaxies as well as ordinary dwarf galaxies with slope $\sim 4$. 
These results hint that LSBGs may owe their properties to the high spins of their host dark matter halos, rather than the halo densities or feedback processes. 
Additionally, our findings do not rule out the Modified Newtonian Dynamics (MOND) theory.
\par

Our findings conflict with the BTFR results of UDGs (\citealp{Pina2019-UDG_BTFR, Guo2020-UDG, Hu2023-UDG_BTFR, Karunakaran2020-UDG, Rong2024-UDG}). This is primarily because that our LSBG sample contains very few UDGs. Indeed, almost all of our LSBGs are classic LSBGs, rather than UDGs with extremely large effective radii.
Therefore, we propose that UDGs may be a distinct galaxy population with different BTFR and mass distributions, compared with classical LSBGs. 
They may also have distinct formation mechanisms that warrant further investigations.
\par

\section*{Acknowledgements}
We express our sincere thanks to the referee for the detailed comments and suggestions. We also thank Xufen Wu, Haochen Jiang, Jingshuo Yang, Shijiang Chen and Yu Chen for the inspiring discussions during this study.
Y.R. acknowledges supports from the NSFC grant 12273037, the CAS Pioneer Hundred Talents Program (Category B), the USTC Research Funds of the Double First-Class Initiative (No.~YD2030002013), and the research grants from the China Manned Space Project (the second-stage CSST science projects: ``Investigation of small-scale structures in galaxies and forecasting of observations'' and ``CSST study on specialized galaxies in ultraviolet and multi-band'').
H.H. is supported by the Fundamental Research Funds for the Central Universities, the CAS Project for Young Scientists in Basic Research Grant No. YSBR-062, the National SKA Program of China No. 2022SKA0110201, and the NSFC grant No. 12033008. 
\par

\section*{Data Availability}

Data are available if requested.



\bibliographystyle{mnras}
\bibliography{RAR_LSBG} 

\bsp	
\label{lastpage}
\end{document}